\documentclass[12pt]{article}

\usepackage{graphicx}
\usepackage{a4}

\usepackage{fancybox}
\usepackage{epsfig}
\usepackage{color}

\usepackage{amsfonts}
\usepackage{mathrsfs}
\usepackage{amssymb}
\usepackage{amsmath}

\usepackage{dcolumn}
\usepackage{bm}

\def\pa{\partial}

\def\ep{\epsilon}

\def\th{\theta}

\def\si{\sigma}

\def\Ga{\Gamma}

\def\Om{\Omega}

\newcommand{\ben}{\begin{equation}}
\newcommand{\een}{\end{equation}}
\newcommand{\bea}{\begin{eqnarray}}
\newcommand{\eea}{\end{eqnarray}}
\newcommand{\ba}{\begin{array}}
\newcommand{\ea}{\end{array}}
\newcommand{\bit}{\begin{itemize}}
\newcommand{\eit}{\end{itemize}}

\newcommand{\vs}[1]{\vspace{#1 mm}}

\newcommand{\dsl}{\pa \kern-0.5em /}

\begin{document}

\topmargin 0pt \oddsidemargin 0mm




\vspace{2mm}

\begin{center}

{\Large \bf Generalized split monopole magnetospheres: the effects
of current sheets}

\vs{10}

 {\large Huiquan Li \footnote{E-mail: lhq@ynao.ac.cn} and Jiancheng Wang}

\vspace{6mm}

{\em

Yunnan Observatories, Chinese Academy of Sciences, \\
650216 Kunming, China

Key Laboratory for the Structure and Evolution of Celestial Objects,
\\ Chinese Academy of Sciences, 650216 Kunming, China

Center for Astronomical Mega-Science, Chinese Academy of Sciences,
\\ 100012 Beijing, China}

\end{center}

\vs{9}

\begin{abstract}

We show that energy should be dissipated or extracted in the current
sheet (CS) of a split magnetosphere deviating from the Michel split
monopole, with the CS heating up or cooling down. But the
electromagnetic energy remains unchanged everywhere. Based on the
de-centered monopole solution generated by symmetry in flat
spacetime, we construct two generalized split monopole
configurations, in which the field lines intersect with the CS at
arbitrary angles. One configuration resembles the outer geometry of
the so-called ``new pulsar magnetosphere model", for which up to
$47\%$ of the spin down energy is transferred to the Joule heating
process in the CS. In the other configuration, we observe that
negative energy is dissipated in the CS, which is usually observed
in magnetospheres on rotating black holes. This means that energy is
extracted simultaneously from the central star and the CS to power
the output Poynting flux at infinity. We interpret the extraction of
energy from the CS as that thermal energy of charged particles in
the CS is transferred to the ordered kinetic energy of these
particles drifting in the force-free (FF) electromagnetic fields.
Hence, the CS is like an ``air conditioner" in the sky, which can
heat up or cool down, depending on the configurations.

\end{abstract}



\section{Introduction}
\label{sec:introduction}

To avoid the appearance of magnetic monopole, the splitting
technique by introducing an equatorial current sheet (CS) is usually
adopted in constructing monopole magnetosphere on a compact object.
In the popular pulsar magnetosphere model at present, the near-star
region is of a dipole structure
\cite{1969ApJ...157..869G,1973ApJ...180..207M}. The magnetic field
lines around the poles can extend beyond the light cylinder (LC) to
infinity. They asymptotically approach a split monopole in the outer
region, whose analytical solution is the one found by Michel
\cite{1973ApJ...180L.133M}. This split monopole results from a
splitting and then gluing of two centered monopoles with opposite
magnetic charge. The discontinuity of the fields across the equator
gives rise to a singular CS. In the Michel split monopole, the
surface of the CS is parallel to the magnetic field lines neighbor
to the equator. So the Lorentz force vanishes and no dissipation
occurs in the CS. This structure is part of the standard pulsar
magnetosphere model first realized numerically by
\cite{1999ApJ...511..351C}.

But this splitting method is not unique to obtain a split monopole.
It is possible that the magnetic field lines are not necessarily
parallel to the infinitely thin CS
\cite{2008JCAP...11..002G,2014ApJ...781...46C,2014MNRAS.445.2500G}.
This modification may cause non-trivial effects. As shown in the
numerical solutions \cite{2008JCAP...11..002G,2014ApJ...781...46C},
the non-parallel splitting leads to a CS where the spin down energy
is dissipated. It is unclear wether this dissipation process
consumes the electromagnetic energy in the CS. If it does (like the
magnetic reconnection case), the magnetosphere should evolve in
time.

In some other numerical simulations on rotating black holes, an
alternative role of the CS is explored. In
\cite{2004MNRAS.350..427K,2005ApJ...620..889U,2017PhRvD..96f3006C,
2018PhRvD..98b3008E}, it is found that the energy dissipated in the
CS developed within the ergosphere is gained by the force-free (FF)
fields to power the jet formation. The origin of the negative
dissipation energy remains vague. It looks like that this phenomenon
is specific to a gravitational system.

In this work, we construct two analytical split monopole models that
the magnetic field lines intersect with the infinitely thin CS at
arbitrary angles. Using these exact configurations, we can clarify
the current and energy flows in the systems in detail, and specify
the precise effects of the CS. The paper is organized as follows. In
terms of the translational freedom mentioned in Section.\
\ref{sec:trans}, we give the general monopole solution whose center
can be shifted along the spin axis in Section.\ \ref{sec:genmonsol}.
In Section.\ \ref{sec:splitmon}, we present the generalized split
monopole configurations based on the de-centered solution and
discuss the effects of the CS by calculating the exact amount of
flows in them. Finally, we summarize and discuss in the last
section.

\section{Translation of the magnetosphere}
\label{sec:trans}

We consider the force-free magnetosphere on an axisymmetric rotator.
The fields satisfy the following FF condition
\begin{equation}\label{e:ffcon}
 \rho\vec{E}+\vec{j}\times\vec{B}=0.
\end{equation}
This condition implies $\vec{j}\cdot\vec{E}=0$ and
$\vec{E}\cdot\vec{B}=0$.

Under this condition, the Maxwell's equations can be reduced to a
simple system described by three correlated functions: the flux
$\psi$, the angular velocity $\Om(\psi)$ of field lines and the
poloidal electric current $I(\psi)$. In terms of them, The
electromagnetic fields in the unit basis of spherical coordinates
can be expressed as
\begin{equation}\label{e:E}
 \vec{E}=-\vec{V}_\phi\times\vec{B}
=-\frac{\Om(\psi)}{r}\left(r\pa_r\psi, \pa_\th\psi, 0\right),
\end{equation}
\begin{equation}\label{e:B}
 \vec{B}=\frac{1}{r^2\sin\th}\left(\pa_\th\psi, -r
\pa_r\psi, rI(\psi)\right).
\end{equation}
where $V_\phi=r\sin\th\Om$. The charge and current densities are
respectively
\begin{equation}\label{e:charge}
 \rho=-\frac{1}{4\pi}\vec{\nabla}\cdot\left(\Om
\vec{\nabla}\psi\right),
\end{equation}
\begin{equation}\label{e:current}
 \vec{j}=\rho r\sin\th\Om\vec{e}_\phi+\frac{1}{4\pi}I'\vec{B}.
\end{equation}

With the above relations and equations, we arrive at the so called
force-free pulsar  magnetosphere equation. By redefining
$z=r\cos\th$ and $x=r\sin\th$, we express the equation in
cylindrical coordinates as
\begin{equation}\label{e:GSeqc}
 (1-\Om^2x^2)\left(\pa_x^2\psi+\pa_z^2\psi\right)
-\frac{1}{x}(1+\Om^2x^2)\pa_x\psi+(\pa_z\psi)^2]=-I(\psi)I'(\psi),
\end{equation}
where the primes stand for derivative with respect to $\psi$.

It is easy to see that the differential equation is invariant under
the shift along the symmetry axis:
\begin{equation}\label{e:trans}
 z\rightarrow z'=z-\ep.
\end{equation}
So any solution $\psi$ shifted along the rotation axis is still a
solution to the pulsar equation (\ref{e:GSeqc}). Under this
translation, the functional relations $\Om(\psi)$, $I(\psi)$ and the
global features are kept the same. In what follows, we consider the
translated version of Michel's monopole solution.

\section{De-centered monopole solution}
\label{sec:genmonsol}

The exact monopole solution found by Michel
\cite{1973ApJ...180L.133M} is quite simple, only relying on the
angle $\th$ in spherical coordinates:
\begin{equation}\label{e:michelsol}
 \psi(\th)=-q\cos\th=-q\frac{z}{\sqrt{x^2+z^2}}.
\end{equation}
For an arbitrary $\Om(\psi)$, the electric current is:
\begin{equation}\label{e:I}
 I(\psi)=\frac{1}{q}\Om(\psi)(\psi^2-q^2),
\end{equation}
where $q$ is the charge of the monopole. The solution gives rise to
a magnetosphere with magnetic domination and null current.

The above solution (\ref{e:michelsol}) under the translation
(\ref{e:trans}) becomes
\begin{equation}\label{e:gensol}
 \psi(r,\th)=-q\frac{z-\ep}{\sqrt{x^2+(z-\ep)^2}}
=-q\frac{r\cos\th-\ep}{\sqrt{r^2-2\ep r\cos\th+\ep^2}}.
\end{equation}
The solution is now dependent on both poloidal coordinates in the
spherical coordinates. The functional relations between $I(\psi)$,
$\Om(\psi)$ and $\psi$ remain invariant, the same as shown in Eq.\
(\ref{e:I}). The solution describes a monopole magnetosphere whose
center is shifted away from the origin (the center of the star)
along the rotation axis by a distance $\ep$, which generalizes the
coincident Michel's solution.

For the solution, the electromagnetic fields are
\begin{equation}\label{e:}
 \vec{E}=\frac{qr\sin\th\Om}{D^3}
\left(\ep\sin\th, -(r-\ep\cos\th), 0\right),
\end{equation}
\begin{equation}\label{e:}
 \vec{B}=\frac{q}{D^3}\left(r-\ep\cos\th,
\ep\sin\th, -r\sin\th\Om D\right).
\end{equation}
where $D=\sqrt{r^2-2\ep r\cos\th+\ep^2}$. Thus, the invariant is
\begin{equation}\label{e:}
 \vec{B}^2-\vec{E}^2=\frac{q^2}{D^4}>0,
\end{equation}
So it is also magnetically dominated.

The relations $B_\phi=E_\th=V_\phi B_r$ for the original Michel's
solution are replaced by
\begin{equation}\label{e:}
 B_\phi=-\sqrt{E_r^2+E_\th^2}=V_\phi\sqrt{B_r^2+B_\th^2}.
\end{equation}
But, at large distances $r\gg|\ep|$, the former is still a good
approximation.

The Poynting flux is
\begin{equation}\label{e:poynting}
 \vec{S}=\frac{1}{4\pi}\vec{E}\times\vec{B}
=\frac{(qr\sin\th\Om)^2}{4\pi D^5}\left(r-\ep\cos\th, \ep\sin\th,
\frac{D}{r\sin\th\Om}\right).
\end{equation}
From this, we can find that the the drift velocity
$\vec{v}_D=4\pi\vec{S}/B^2$ gets a non-vanishing component at the
$\th$ direction. The four-current is
\begin{equation}\label{e:4current}
 J^\mu\equiv(\rho,\vec{j})=-\frac{q\Om(r\cos\th-\ep)}{2\pi D^4}
\left(D, r-\ep\cos\th, \ep\sin\th, 0\right).
\end{equation}
So it is null with $J^2=0$, meaning the particle travels at speed of
light. This is the same as Michel's centered solution. But the null
surface where the charge density and current vanish is not located
on the equator any more, but on the plane shifted by a distance
$\ep$: $z=r\cos\th=\ep$. It is noticed that the poloidal components
of the magnetic field, Poynting flux, the drift velocity and the
current are all parallel to each other.

\section{Generalized split monopoles}
\label{sec:splitmon}

Since the magnetic monopole has not yet been confirmed, the
splitting technique is usually adopted in constructing a monopole
magnetosphere. For the Michel solution, the split monopole
configuration on the two half-planes is expressed as:
\begin{equation}\label{e:censplitmon}
 \psi(\th)=
 \left\{
 \begin{array}{cl}
 q(1-\cos\th),
\textrm{ }\textrm{ }\textrm{
} \th\in[0,\pi/2) \\
 q(1+\cos\th).
\textrm{ }\textrm{ }\textrm{ } \th\in(\pi/2,\pi]
 \end{array}
 \right.
\end{equation}
The splitting results in discontinuity on the equatorial plane,
giving rise to an infinitely thin CS there. In this case, the
magnetic field lines are parallel to the surface of the CS.

With the de-centered solution (\ref{e:gensol}), we make the similar
splitting. We denote the solution on the upper half-plane by
$\psi_\vee$ and the one on the lower hemisphere by $\psi_\vee$. On
the upper hemisphere $\th\in[0,\pi/2)$, we choose $\ep=\pm d$
($d>0$) and express the solution as
\begin{equation}\label{e:}
 \psi_\vee^{(\pm)}=q\left(1-\frac{r\cos\th\pm d}
{\sqrt{r^2\pm 2dr\cos\th+d^2}}\right).
\end{equation}
On the lower one $\th\in(\pi/2,\pi]$, the solution is
\begin{equation}\label{e:}
 \psi_\wedge^{(\pm)}=q\left(1+\frac{r\cos\th\pm d}
{\sqrt{r^2\pm2dr\cos\th+d^2}}\right).
\end{equation}

There are two trivial cases:
$\psi=(\psi_\vee^{(+)},\psi_\wedge^{(+)})$ and
$\psi=(\psi_\vee^{(-)},\psi_\wedge^{(-)})$, which just describe
de-centered versions of the Michel split monopole magnetosphere,
shifted as a whole respectively downward and upward by a distance
$d$. We are more interested in the two non-trivial configurations
that will be discussed as follows.

\begin{figure}
    \center
    \includegraphics[width=0.49\columnwidth]{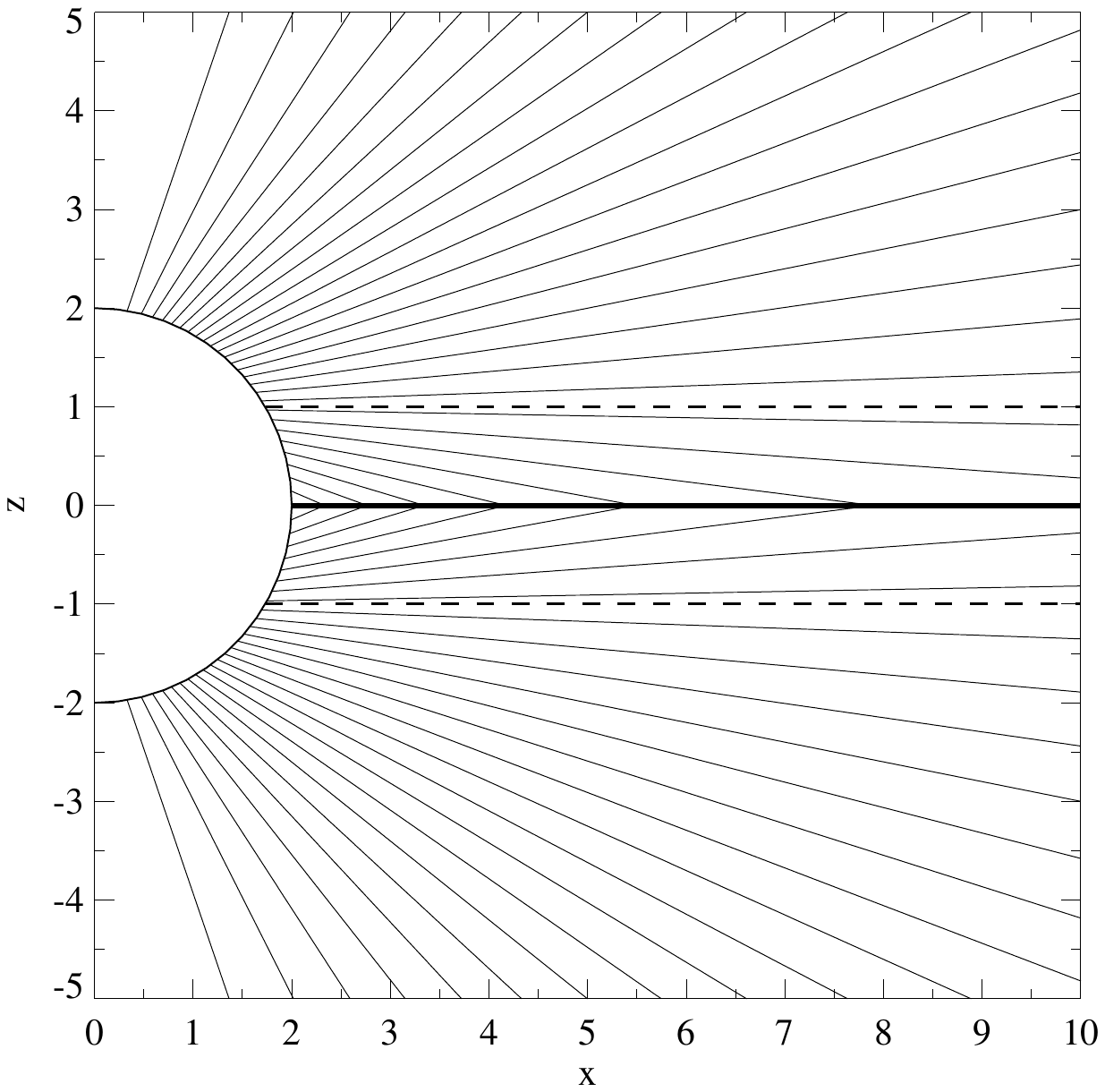}
    \includegraphics[width=0.49\columnwidth]{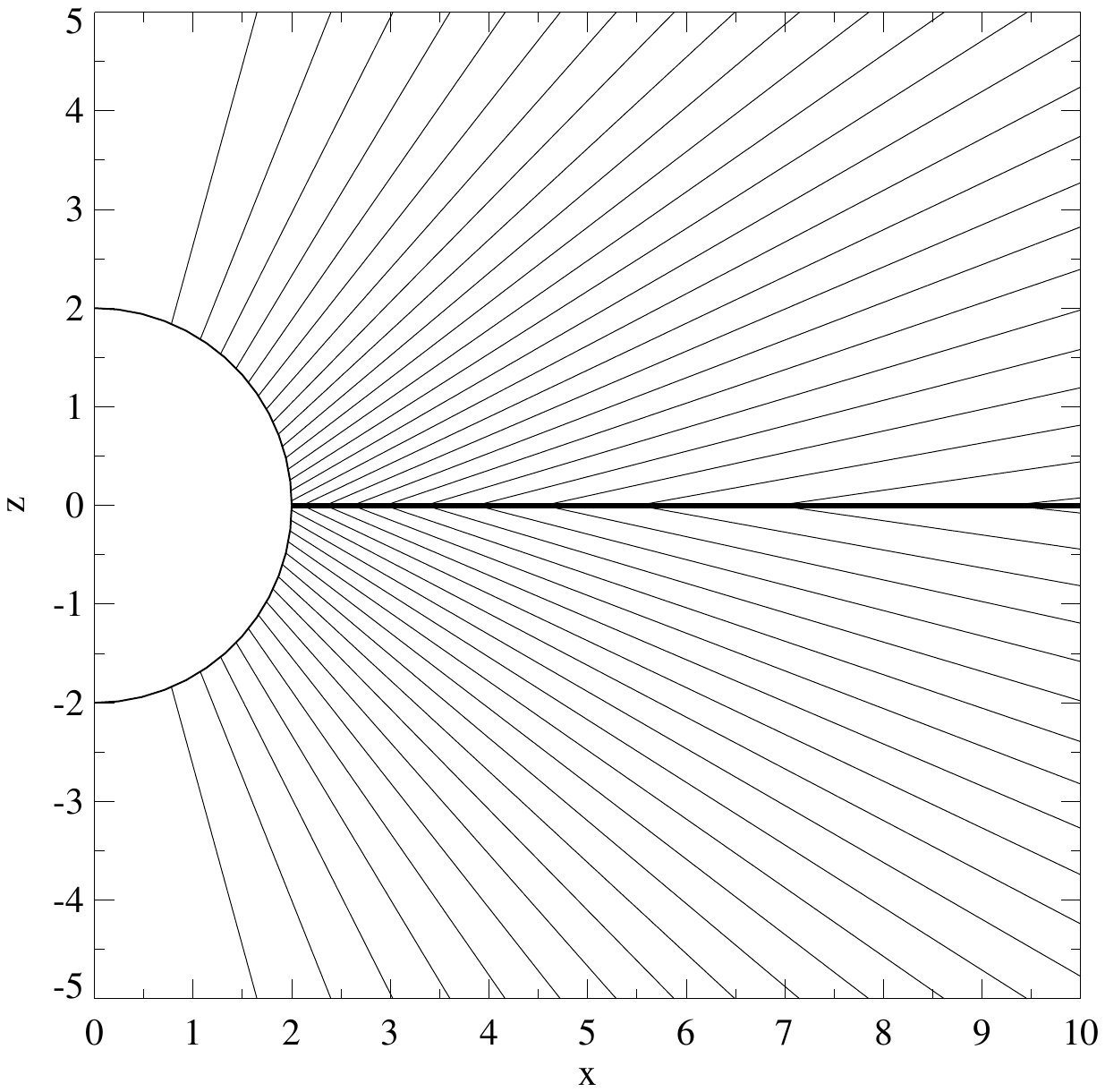}
    \caption{Magnetic field lines for the configurations defined in
    Section.\ \ref{subsec:con_I} (left) and \ref{subsec:con_II} (right) with
    $q=1$, $r_0=2$ and $d=1$. The bold lines on the equators
    represent the infinitely thin CS. The dashed lines on the
    left panel represent the null surfaces (with $\rho=0$), across which
    the charge density $\rho$ changes sign. In a realistic pulsar
    magnetosphere model, these split monopole configurations are relevant
    only to the geometry outside the LC.}
    \label{fig:f1}
\end{figure}

\subsection{$\psi=(\psi_\vee^{(-)},\psi_\wedge^{(+)})$}
\label{subsec:con_I}

This configuration is shown in the left panel of Fig.\ \ref{fig:f1}.
The profile of the configuration resembles the outer geometry in the
``new pulsar magnetosphere model" obtained numerically in
\cite{2014ApJ...781...46C}. The valid region of the configuration is
restricted to be $r>d$, where integral of the magnetic field over
any closed surface leads to zero magnetic charge.

The expanded form of the split solution can be obtained in terms of
the general expanded form in the Appendix. It is clearly seen that
the full expanded solution of this spit monopole in the outer range
$d/r<1$ is a summation of a closed and an open field line component:
the odd order terms of $r$ are continuous across the equatorial
plane, contributing a closed field line component, while the even
order terms are discontinuous, contributing an open field line
component. The discontinuity in the latter gives rise to a current
sheet on the equatorial plane.

For $d/r<1$, the solution expanded to the order
$\mathcal{O}(d^2/r^2)$ is
\begin{equation}\label{e:expsol}
 \psi\simeq
 \left\{
 \begin{array}{cl}
 q[1-\cos\th+d\sin^2\th r^{-1}+(3d^2/2)\cos\th\sin^2\th r^{-2}],
\textrm{ }\textrm{ }\textrm{
} \th\in[0,\pi/2) \\
 q[1+\cos\th+d\sin^2\th r^{-1}-(3d^2/2)\cos\th\sin^2\th r^{-2}].
\textrm{ }\textrm{ }\textrm{ } \th\in(\pi/2,\pi]
 \end{array}
 \right.
\end{equation}
At infinity $r\rightarrow\infty$, the solution is asymptotically the
Michel split monopole solution. In the near region with small $d/r$
(but not too small), the dipole part becomes more important.
Besides, there are two null surfaces in this split monopole
solution. This is coincident to the case of the corotating dipole
inside the LC in the standard pulsar magnetosphere model, where two
straight null surfaces extend from the star surface to the LC.
Hence, the new split solution here can better describe a smooth
transition from a dipole to a monopole. By adjusting the parameter
$d$, the two null lines in corotating dipole and in new split
monopole can be matched. In particular, the expanded solution
(\ref{e:expsol}) is exactly the solution outside the light torus of
the exact dipole magnetosphere \cite{2016arXiv160807998P}.

Let us now examine the dynamical consequence in this configuration.
From the solution, the force-free fields approaching the equator
$\th\rightarrow\pi/2$ from either side are given by
\begin{equation}\label{e:}
 \vec{E}_\vee\rightarrow\frac{qr\Om}{(r^2+d^2)^
{\frac{3}{2}}}\left(d,-r,0\right),
\end{equation}
\begin{equation}\label{e:}
 \vec{E}_\wedge\rightarrow\frac{qr\Om}{(r^2+d^2)^
{\frac{3}{2}}}\left(d,r,0\right).
\end{equation}
\begin{equation}\label{e:}
 \vec{B}_\vee\rightarrow\frac{q}{(r^2+d^2)^
{\frac{3}{2}}}\left(r,d,-r\Om\sqrt{r^2+d^2}\right),
\end{equation}
\begin{equation}\label{e:}
 \vec{B}_\wedge\rightarrow\frac{q}{(r^2+d^2)^
{\frac{3}{2}}}\left(-r,d,r\Om\sqrt{r^2+d^2}\right).
\end{equation}
We denote the continuous fields at the equator as:
\begin{equation}\label{e:CSfield}
 E_c^r\equiv E^r_\vee(\th\rightarrow\frac{\pi}{2})
=E^r_\wedge(\th\rightarrow\frac{\pi}{2}), \textrm{ }\textrm{
}\textrm{ }
B_c^\th\equiv B^\th_\vee(\th\rightarrow\frac{\pi}{2})
=B^\th_\wedge(\th\rightarrow\frac{\pi}{2}).
\end{equation}
They are the field components that are non-vanishing within the CS.
The other field components are discontinuous.

The discontinuity of the perpendicular electric field $E^\th$ leads
to the surface charge density in the CS:
\begin{equation}\label{e:}
 \si_c=\frac{qr^2\Om}{2\pi(r^2+d^2)^{\frac{3}{2}}}.
\end{equation}
The discontinuities of the parallel magnetic fields $B^r$ and
$B^\phi$ give rise to the surface current densities flowing in the
CS respectively along the $r$ and $\phi$ directions:
\begin{equation}\label{e:CScurrent}
 i_c^\phi=\frac{qr}{2\pi(r^2+d^2)^{\frac{3}{2}}},
\textrm{ }\textrm{ }\textrm{ } i_c^r=\frac{qr\Om}{2\pi(r^2+d^2)}.
\end{equation}

In the FF regions, the total change rate of the charges through the
sphere (excluding the equator) at radius $r$ is
\begin{equation}\label{e:}
 \dot{Q}^{FF}(r)=\int_\vee j^r_\vee ds^r+\int_\wedge j^r_\wedge ds^r
=\left[I_\vee(\psi)\right]_{\th=0}^{\th=\pi/2}
+\left[I_\wedge(\psi)\right]^{\th=\pi}_{\th=\pi/2},
\end{equation}
where $ds^r=2\pi r^2\sin\th d\th$ and the dot denotes the derivative
with respect to time. The change rate through the section of the CS
at $r$ is
\begin{equation}\label{e:}
 \dot{Q}^{CS}(r)=2\pi ri_c^r.
\end{equation}
Thus, it is justified that the total electric current flowing
through a sphere at any radius $r$ is zero:
\begin{equation}\label{e:neutralcon}
 \dot{Q}^{FF}(r)+\dot{Q}^{CS}(r)=0.
\end{equation}
This implies that the central star always remains neutral.

We take the value $\dot{Q}^{CS}(r=r_0)$ at some initial radius $r_0$
$(>d)$ as the current directly from the central star and the one
$\dot{Q}^{CS}(r\rightarrow\infty)$ at infinity as the output
current. From the second equation of Eq.\ (\ref{e:CScurrent}), the
latter is given by
\begin{equation}\label{e:}
 \dot{Q}^{CS}(r\rightarrow\infty)=q\Om.
\end{equation}
It is the same as the Michel split monopole.

Towards the equator, the perpendicular electric currents along the
FF magnetic fields are
\begin{equation}\label{e:}
 j^\th_\vee (\th\rightarrow\frac{\pi}{2})=-j^\th_\wedge
(\th\rightarrow\frac{\pi}{2})=\frac{q\Om d^2}{2\pi(r^2+d^2)^2}.
\end{equation}
This means that there is a net electric current flowing into the CS
from both the upper and the lower sides. Including the injected
current at $r_0$ and the output current at $r\rightarrow\infty$, we
find
\begin{equation}\label{e:currentcon}
 \dot{Q}^{CS}(r=r_0)
+\int_{r_0}^\infty 2\pi r\left[j^\th_\vee(\th\rightarrow
\frac{\pi}{2})-j^\th_\wedge(\th\rightarrow\frac{\pi}{2})\right]dr
=\dot{Q}^{CS}(r\rightarrow\infty).
\end{equation}
This equation says that the output CS current comes from directly
the central star and the FF fields.

With the continuous fields, we obtain the non-vanishing Lorentz
force densities in the CS:
\begin{equation}\label{e:}
 f_c^r=\si_cE_c^r-i_c^\phi B_c^\th
=\frac{q^2rd(r^2\Om^2-1)}{2\pi(r^2+d^2)^3},
\end{equation}
\begin{equation}\label{e:}
 f_c^\phi=i_c^r B_c^\th
=\frac{q^2rd\Om}{2\pi(r^2+d^2)^{\frac{5}{2}}}.
\end{equation}
Both tend to zero in the Michel split monopole solution with $d=0$.
The directions of the radial component are opposite on the two sides
of the LC located at $r=r_{LC}=1/\Om$: the magnetic force dominates
inside the LC, while the electric force dominates outside the LC.
The fields become electrically dominated outside LC either. It is
usually assumed that the monopole solution only exists outside the
LC.

\begin{figure}[htbp]
    \center
    \includegraphics[width=0.95\columnwidth]{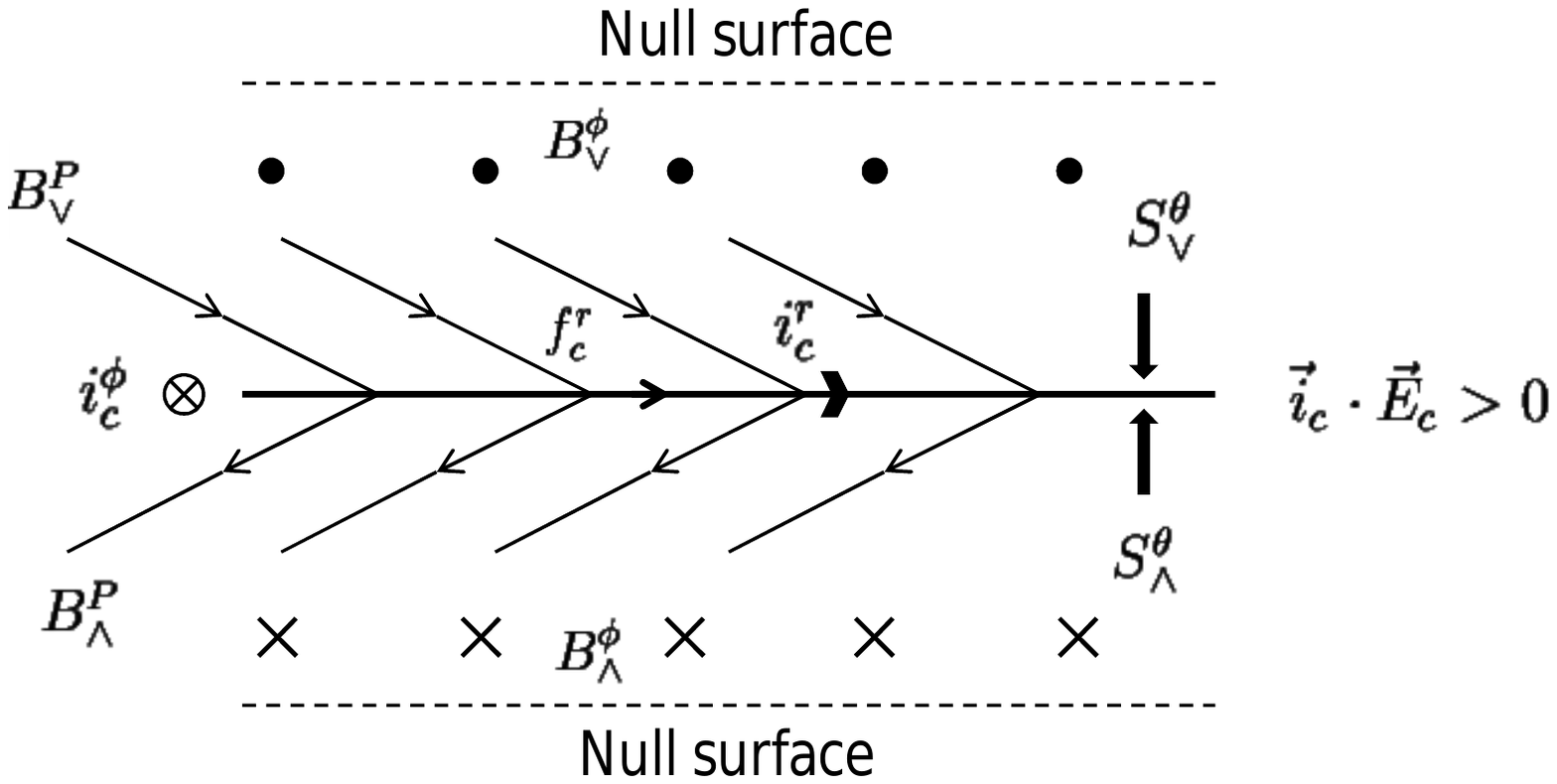}
    \caption{The field structure and the energy flows near the CS
    outside the LC of the configuration in Section.\ \ref{subsec:con_I}.
    The electric field lines (not displayed) are perpendicular to the
    poloidal magnetic field lines $B^P$. A net Poynting flux flows
    into the CS from the upper and lower sides to provide the energy
    dissipated in the CS.}
    \label{fig:f2}
\end{figure}

The Poynting fluxes for the FF fields in the limit
$\th\rightarrow\pi/2$ are:
\begin{equation}\label{e:}
 \vec{S}_\vee\rightarrow\frac{(qr\Om)^2}{4\pi(r^2+d^2)^
{\frac{5}{2}}}\left(r,d,\frac{1}{r\Om}\sqrt{r^2+d^2}\right),
\end{equation}
\begin{equation}\label{e:}
 \vec{S}_\wedge\rightarrow\frac{(qr\Om)^2}{4\pi(r^2+d^2)^
{\frac{5}{2}}}\left(r,-d,\frac{1}{r\Om}\sqrt{r^2+d^2}\right).
\end{equation}
The discontinuous perpendicular component indicates that there is a
net Poynting flux flowing into the CS from both sides of the FF
regions. Thus, the CS gains energy.

The FF regions should be dissipation free and the electromagnetic
energy should always be conserved. This can be expressed as:
\begin{equation}\label{e:encon}
 \dot{\mathcal{E}}^{FF}_\vee=\dot{\mathcal{E}}^{FF}_\wedge=0,
\end{equation}
where the change rates of the electromagnetic energy are given by
\begin{equation}\label{e:Poyntingint}
 \dot{\mathcal{E}}^{FF}_\vee=\oint_\vee\vec{S}_\vee\cdot d\vec{s},
\textrm{ }\textrm{ }\textrm{ }
\dot{\mathcal{E}}^{FF}_\wedge=\oint_\wedge\vec{S}_\wedge \cdot
d\vec{s}.
\end{equation}
The integrals go through all the boundaries of the FF regions.

We first consider the upper hemisphere. The change rate of the FF
electromagnetic energy due to the Poynting influx crossing the
hemisphere at the initial radius $r_0$ is
\begin{equation}\label{e:}
 \dot{\mathcal{E}}^{FF}_\vee(r=r_0)=2\pi r_0^2\int_{0}^{\frac{\pi}{2}}
\sin\th S^r_\vee d\th=\frac{q^2\Om^2}{6}\left[
2+\frac{d(3r_0^2+2d^2)}{(r_0^2+d^2)^{\frac{3}{2}}}\right].
\end{equation}
This influx can be viewed as the one directly extracted from the
central star (via the inner magnetosphere). The change rate measured
at $r\rightarrow\infty$:
\begin{equation}\label{e:}
 \dot{\mathcal{E}}^{FF}_\vee(r\rightarrow\infty)=-\frac{q^2\Om^2}{3}.
\end{equation}
The negative sign means that the energy flows out the FF region to
infinity. This result is also the same as the Michel solution. On
the boundary along the equator, the energy also flows out the FF
region:
\begin{equation}\label{e:}
 \dot{\mathcal{E}}^{FF}_\vee(\th\rightarrow\frac{\pi}{2})
=-\int_{r_0}^{\infty} 2\pi r S_\vee^\th
(\th\rightarrow\frac{\pi}{2})dr=
-\frac{dq^2\Om^2(3r_0^2+2d^2)}{6(r_0^2+d^2)^{\frac{3}{2}}},
\end{equation}

The calculations on the lower hemisphere lead to identical results:
$\dot{\mathcal{E}}^{FF}_\wedge=\dot{\mathcal{E}}^{FF}_\vee$ for each
of the components. We denote the summation:
$\dot{\mathcal{E}}^{FF}=\dot{\mathcal{E}}^{FF}_\vee+
\dot{\mathcal{E}}^{FF}_\wedge=2\dot{\mathcal{E}}^{FF}_\vee$. Then we
have totally
\begin{equation}\label{e:}
 \dot{\mathcal{E}}^{FF}(r=r_0)+\dot{\mathcal{E}}^{FF}
(\th\rightarrow\frac{\pi}{2})
+\dot{\mathcal{E}}^{FF}(r\rightarrow\infty)=0.
\end{equation}
Thus, the conservation law (\ref{e:encon}) is verified. This
indicates that the energy extracted from the star sources the
Poynting fluxes flowing into the CS and to infinity. For the Michel
split monopole, the second term vanishes and so the Poynting flux is
constant through any sphere.

We now turn to the energy conservation law in the CS. As given in
Eq.\ (\ref{e:CSfield}), the fields exist inside the CS are the
continuous fields: $E_c^r$ and $B_c^\th$. They give a toroidal
Poynting flux inside the CS, which is conserved itself. The electric
current flowing in and out the CS is also conserved in terms of Eq.\
(\ref{e:currentcon}). So the only change of the energy in the CS
comes from the Poyinting influx $-\dot{\mathcal{E}}^{FF}
(\th\rightarrow\pi/2)$ and the dissipated energy. The latter arises
from the Joule heating process due to the non-vanishing
$\vec{i}_c\cdot\vec{E}_c$. It leads to an increase of the CS energy
at a total rate:
\begin{equation}\label{e:work}
 \dot{\mathcal{E}}^{CS}=\int_{r_0}^{\infty}2\pi r i_c^rE_c^rdr.
\end{equation}
It is clear that this dissipated energy is completely compensated by
the Poynting influx from both sides of the FF fields:
$i_c^rE_c^r=2S_\vee^\th (\th\rightarrow\pi/2)$ or
\begin{equation}\label{e:CSeqFF}
 -\dot{\mathcal{E}}^{FF}(\th\rightarrow\frac{\pi}{2})
=\dot{\mathcal{E}}^{CS}.
\end{equation}
Hence, the energy is conserved and there is also no electromagnetic
energy lost in the CS.

The equation (\ref{e:CSeqFF}) should be a consequence of the
following process: as the charged particles flow into the CS along
the magnetic field lines, the perpendicular component of the drift
velocity will be eventually damped to zero. The kinetic energy is
transferred to the thermal internal energy in the CS (as shown in
Figure.\ \ref{fig:f2}).

Compared with the Michel split monopole, the spin down power is
enhanced in this configuration. The ratio of dissipated energy to
the total extracted energy is
\begin{equation}\label{e:}
 \frac{\dot{\mathcal{E}}^{CS}}{\dot{\mathcal{E}}^{FF}(r=r_0)}=
\left[1+\frac{2(r_0^2+d^2)^\frac{3}{2}}{d(3r_0^2+2d^2)}\right]^{-1}.
\end{equation}
Since $r_0>d$, the maximum energy that can be dissipated in the CS
is $47\%$ of the total spin down energy, which is close to the
numerical result of \cite{2014ApJ...781...46C}. For larger
$r_0/d>1$, a smaller portion of energy is dissipated.

\subsection{$\psi=(\psi_\vee^{(+)},\psi_\wedge^{(-)})$}
\label{subsec:con_II}

The magnetic field distribution of this configuration is shown in
the right panel of Fig.\ \ref{fig:f1}. It looks similar to the split
magnetosphere in the presence of a thin accretion disk that contains
magnetic fields itself (e.g.,
\cite{1981MNRAS.196.1021B,2004MNRAS.350..427K,2005ApJ...620..889U,
2017PhRvD..96f3006C,2020arXiv201014470C}). But here the CS is not an
accretion disk since no gravity is involved.

The quantities for this configuration is given by the previous case
just with the replacement $d\rightarrow-d$. In the CS, only the
continuous fields $E_c^r$ and $B_c^\th$ exist. The discontinuous
fields lead to surface charge density $\si_c$ and current densities
$i_c^\phi$, $i_c^r$, which are the same as the previous case. The
currents also close with the same forms as given in Eqs.\
(\ref{e:neutralcon}) and (\ref{e:currentcon}). But the Lorentz force
take the opposite directions:
\begin{equation}\label{e:}
 f_c^r=-\frac{q^2rd(r^2\Om^2-1)}{2\pi(r^2+d^2)^3},
\textrm{ }\textrm{ }\textrm{ }
f_c^\phi=-\frac{q^2rd\Om}{2\pi(r^2+d^2)^{\frac{5}{2}}}.
\end{equation}

The Poynting fluxes perpendicular to the CS are:
\begin{equation}\label{e:}
 S_\vee^\th(\th\rightarrow\frac{\pi}{2})=-S_\wedge^\th
(\th\rightarrow\frac{\pi}{2})\rightarrow-
\frac{d(qr\Om)^2}{4\pi(r^2+d^2)^{\frac{5}{2}}}.
\end{equation}
It indicates that net Poynting fluxes flow off the CS into the FF
magnetosphere on both sides. So the FF magnetosphere gains energy
from the CS. Integrating the Poynting flux along the equator, we can
find that the energy gained by the FF fields is exactly that lost in
the CS:
\begin{equation}\label{e:CSconII}
 \dot{\mathcal{E}}^{FF}(\th\rightarrow\frac{\pi}{2})
=-\dot{\mathcal{E}}^{CS}
=\frac{dq^2\Om^2(3r_0^2+2d^2)}{3(r_0^2+d^2)^{\frac{3}{2}}}.
\end{equation}
So the energy is conserved in the CS and the electromagnetic energy
density remains unchanged.

\begin{figure}[htbp]
    \center
    \includegraphics[width=0.95\columnwidth]{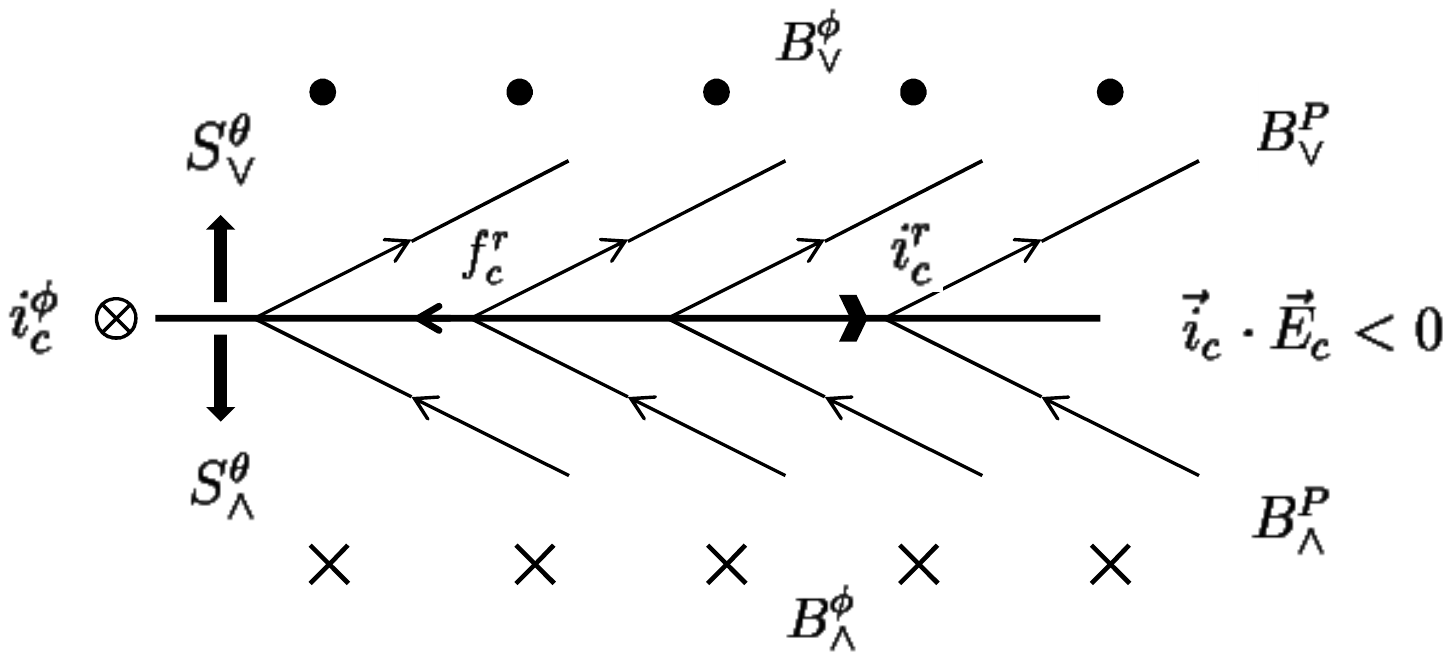}
    \caption{The field structure and the energy flows near the CS
    outside the LC of the configuration in Section.\
    \ref{subsec:con_II}. A net Poynting flux flows off the CS
    to the FF regions on both sides as the CS loses energy.}
    \label{fig:f3}
\end{figure}

Similarly, we can show that the electromagnetic energy is conserved
in the FF regions. With the above equation (\ref{e:CSconII}), the
conservation law can be expressed as:
\begin{equation}\label{e:}
 \dot{\mathcal{E}}^{FF}(r=r_0)-\dot{\mathcal{E}}^{CS}=
\dot{\mathcal{E}}^{FF}(r\rightarrow\infty),
\end{equation}
where $\dot{\mathcal{E}}^{FF}(r\rightarrow\infty)$ is the same as
the previous case, also equal to the one in the Michel split
monopole. This equation means that the output energy flux at
infinity is simultaneously extracted from the central star and the
CS. For a given output power, the spin down energy extracted from
the star can only be $11.6\%$ of that by the Michel split monopole
since $r_0/d>1$.

Notice that here $\dot{\mathcal{E}}^{CS}$ is negative since
$\vec{i}_c\cdot\vec{E}_c=i^r_c E^r_c<0$. This mysterious negative
energy has been encountered in the numerical simulations on rotating
black holes
\cite{2004MNRAS.350..427K,2017PhRvD..96f3006C,2018PhRvD..98b3008E}.
It may be due to the observational effect in the gravitational
system. But here no gravity is involved in our system, which may
bring us new understanding on it. We think that the negative energy
here arises from the observational effect in the co-rotation frame,
which is an acceleration frame and takes similarity to a
gravitational system in terms of the Einstein equivalent principle
between gravity and acceleration.

Following the above analysis, we can interpret this negative energy
process as an inverse process of the one discussed in the previous
configuration (See Figure.\ \ref{fig:f3}): the charges flow away
from the CS to the FF regions along the magnetic field lines on both
sides to form the electric currents that constitute the split
monopole configuration. Then the internal energy of thermal motion
of the particles in the CS is transferred into the ordered drift
motion when the particles enter into the FF regions. So the CS
should cool down with the negative energy dissipated to provide the
extracted energy.

\section{Conclusions and discussions}
\label{sec:con}

The Michel split monopole model is not unique and the deviation from
it leads to non-trivial consequence. By varying the centered model
in different ways, we illustrate how the CS plays different roles.

Based on the de-centered monopole solution generated by the
translational symmetry in the axisymmetric case, we construct two
generalized split monopole configurations. One configuration
resembles the outer geometry of a new pulsar magnetosphere model,
while the other may be useful in describing the physical process in
a split magnetosphere with an accretion disk. These generalized
configurations can also be constructed in the oblique rotation case,
since the translational symmetry still exists in the magnetosphere
on an oblique rotator \cite{1998MNRAS.297..315U}.

It is shown that the CS is a site where energy is dissipated or
extracted. This will increase or decrease the spin down energy
extracted from the central star, for given output Poynting flux. We
interpret this process as a result that the internal energy of
thermal motion and the kinetic energy of drift motion are
transferred into each other. The electromagnetic energy is always
not lost everywhere, i.e., in the CS and the FF regions. When the
Poynting flux flows in, the CS is heated up and possibly leads to
synchrotron and inverse Compton radiations, which are observable
\cite{2016JPlPh..82e6302P}. On the contrary, energy is extracted as
the CS cools down. So the CS can also cause temperature
discontinuities in the systems. The effects of the thermal
non-equilibrium on the magnetohydrodynamics need further
investigations.

Our results will also apply to any variation of the split monopole
in the standard pulsar magnetosphere model. In a realistic
situation, the split monopole should not be exactly like the Michel
model. The CS may have finite size, different geometries or even be
dynamical with wavy structures. So all these variations will cause
extra energy dissipation or extraction in the CS in terms of our
results above.

\section*{Acknowledgements\markboth{Acknowledgements}{Acknowledgements}}

This work is supported by the Yunnan Natural Science Foundation
2017FB005.


\appendix

\section*{Appendix}

\setcounter{equation}{0}

\section{Expansions of the de-centered monopole solution}

In this appendix, we present the expanded forms of the de-centered
monopole solution
\begin{equation}\label{e:gensola}
 \psi(r,\th)=-q\frac{r\cos\th-\ep}{\sqrt{r^2
-2\ep r\cos\th+\ep^2}}=-q\frac{\vec{\ep}}{|\vec{\ep}|}
\cdot\frac{\vec{r}-\vec{\ep}}{|\vec{r}-\vec{\ep}|},
\end{equation}
where $\vec{\ep}$ is a constant vector on the axis. This can be done
by using the generating function for the Legendre polynomials:
\begin{equation}\label{e:}
 \frac{1}{\sqrt{1-2xt+t^2}}=\sum_{n=0}^\infty P_n(x)t^n.
\end{equation}

Let us first consider the expansions in the region $r>|\ep|$. Using
the following identity
\begin{equation}\label{e:}
 \frac{1}{n+1}\sin\th P_{n+1}^1(x)=xP_{n+1}(x)-P_n(x),
\end{equation}
we obtain the expansion form
\begin{equation}\label{e:}
 \psi(r,\th)=\psi_0+\sum_{n=1}^\infty \psi_{-n}(\th)r^{-n},
\end{equation}
where
\begin{equation}\label{e:}
 \psi_0=-q\cos\th, \textrm{ }\textrm{ }\textrm{ }
\psi_{-n}=-\frac{1}{n}q\ep^n\sin\th P_n^1(\cos\th).
\end{equation}

Using the identity
\begin{equation}\label{e:}
 \frac{1}{n+1}\sin\th P_n^1(x)=P_{n+1}(x)-xP_n(x),
\end{equation}
we have for $r<|\ep|$
\begin{equation}\label{e:}
 \psi(r,\th)=q+\sum_{n=2}^\infty \psi_{n}(\th)r^{n},
\end{equation}
where
\begin{equation}\label{e:}
 \psi_{n}=\frac{1}{n}q\ep^{-n}\sin\th P_{n-1}^1(\cos\th).
\end{equation}
This branch of expansions is irrelevant in the discussions here.

With the expanded forms, it is easy to find that the generalised
solution (\ref{e:gensola}) can be obtained from the pulsar equation
by adopting the expansion method in \cite{2017PhRvD..96b3014L}. In
doing so, it is interesting to notice a cubic order identity for the
Legendre polynomials that is not yet found elsewhere:
\begin{equation}\label{e:}
 \sin\th\cos\th\pa_\th\Ga_k+(1-k\sin^2\th)\Ga_k=
\sum_{i=0}^k\sum_{j=0}^i\Ga_{k-i}\Ga_{i-j}\Ga_{j},
\end{equation}
where
\begin{equation}\label{e:}
 \Ga_i=\ep^i [P_{i-1}(\cos\th)-\cos\th P_i(\cos\th)],
\textrm{ }\textrm{ }\textrm{ } (i\geq0)
\end{equation}
with the definition $P_l=0$ for negative $l$.

\bibliography{b}
\bibliographystyle{unsrt}

\end{document}